\input{aipcheck}
\documentclass[final]{aipproc}
\usepackage{color}
\usepackage{amsmath,amssymb}
\usepackage{pslatex}
\layoutstyle{6x9}

\begin{document}
\title{S-wave meson scattering up to 2 GeV and its spectroscopy}


\author{M. Albaladejo}{
  address={Departamento de F\'isica, Universidad de Murcia, E-30071, Murcia, Spain}
}

\author{J. A. Oller}{
  address={Departamento de F\'isica, Universidad de Murcia, E-30071, Murcia, Spain}
}

\author{C. Piqueras}{
  address={Departamento de F\'isica, Universidad de Murcia, E-30071, Murcia, Spain}
}

\begin{abstract}
We have performed a thorough study of the meson-meson S-waves with isospin ($I$) 0 and 1/2, up to $\sqrt{s} \simeq 2\ \mathrm{GeV}$. This is the first study that includes 13 channels that have their threshold below that energy. All the resonances below 2 GeV, namely the $f_0(600)$ or $\sigma$, $f_0(980)$, $f_0(1370)$, $f_0(1500)$, $f_0(1710)$ and $f_0(1790)$ for $I=0$, and the $K^*_0(800)$ or $\kappa$, $K^*_0(1430)$ and $K^*_0(1950)$ for $I=1/2$, are generated. We can then extract a clear picture of the spectroscopy, finding that the $f_0(1710)$, together with an important contribution to the $f_0(1500)$, are glueballs. Another pole, which corresponds mainly to the $f_0(1370)$, is a pure octet $I=0$ state, and does not mix with the glueball.
\end{abstract}

\maketitle

\section{Introduction}
The scalar dynamics is a complicated one due to the large number of resonances and
coupled channels that involves. In addition, some of these resonances are very
broad, overlap between each other and are very sensitive to the coupled channels
involved. Another interesting topic is the study of the nature of these
resonances which, in many cases, goes beyond the simple $q\bar{q}$ picture. E.g., one can find
in addition dynamically generated resonances, glueballs, etc. 
 All these reasons motivate our study \cite{Albaladejo:2008qa}, on which we briefly report here,
 of the $I=0$ meson-meson S-wave in terms of 13 coupled channels, namely $\pi\pi$, $K\bar{K}$, $\eta\eta$, $\eta\eta'$, $\eta'\eta'$, $\sigma\sigma$, 
 $\rho\rho$, $\omega\omega$, $K^*\bar{K^*}$, $\omega\phi$, $\phi\phi$, $a_1(1260)\pi$ and $\pi^*(1300)\pi$. Simultaneously, we study the S-wave of $K^-\pi^+$ (involving $I=1/2$ and $3/2$) with the coupled channel scheme, including $K\pi$, $K\eta$ and $K\eta'$. 

\section{Formalism}

To calculate our scattering amplitudes, we use the lowest order $SU(3)$ Chiral Perturbation Theory Lagrangian, $\mathcal{L}_2$, and the lowest order interaction chiral Lagrangian of an octet and singlet of $0^{++}$ resonances, $\mathcal{L}_S$ \cite{pich}.
 The $\pi$, $K$ and $\eta$ form the octet of the lightest pseudoscalar Goldstone bosons. 
However, when considering higher energy regions, as we do here, one has to take into 
account additionally the $\eta\eta'$ and $\eta'\eta'$ channels. In the large $N_c$ limit, the $\eta_1$ becomes the ninth Goldstone boson. This fact can be used to build chiral Lagrangians based on $U(3)$ symmetry rather than on $SU(3)$, including then the $\eta_1$ field. It is well known that
the $\eta_1$ and $\eta_8$ mix to give the physical $\eta$ and $\eta'$ mesons,
 and we take for the mixing angle the value $\sin\theta = -1/3$.

 The  matrix $\Phi=\sum_{i=1}^8 \phi_i\lambda_i/\sqrt{2}+\eta_1/\sqrt{3}$  
  incorporates in a standard way the nonet of the lightest pseudoscalars.
We also employ the matrix  $U=\exp(i \sqrt{2} \Phi/f)$ and the covariant
derivative $D_\mu U=\partial_\mu U -i r_\mu U+i  U \ell_\mu$, with
 $f$ the pion decay constant in the chiral limit fixed to $f_\pi=92.4$~MeV.
  The classical external left and right fields, respectively, $l_{\mu}$ and $r_{\mu}$, are needed to gauge the global chiral symmetry to a local one. We make the identification $v_{\mu} \equiv (r_{\mu} + l_{\mu})/2 = \lambda W_{\mu}$, where $W_{\mu}$ is the nonet of the lightest $1^{--}$ vector resonances (including $\rho$, $K^*$, $\omega$ and $\phi$), and $\lambda = 4.3$ from the $\rho\to\pi\pi$ width.

The interaction kernels, $N_{i,j}$ (the subscripts $i,j$ represent here the channels), 
are calculated from the sum of the Lagrangians $\mathcal{L}_2 + \mathcal{L}_S$. ${\cal L}_2$ corresponds
to local interactions while $\mathcal{L}_S$ gives rise to the $s$-channel exchange of $SU(3)$ multiplets of \textit{bare} resonances.\footnote{Explicit expressions for the simplified situation of just three channels without the 
$\eta_1$ field  can be found in ref. \cite{Oller:1998zr}.} 
 We employ  the master formula $T = (1+N\cdot g)^{-1}\cdot N$ \cite{Oller:1998zr}, 
 where $T$ is a $13\times 13$ matrix that contains the elements $T_{i,j}$ 
 and $g$ is a diagonal matrix with elements $g_i(s)$. The latter are loop functions 
  which represent the two meson $s-$ channel unitarity loop and satisfy a once subtracted dispersion relation \cite{Oller:1998zr}. The previous equation embodies the coupled channel interactions driven by the 
  $N_{i,j}$ kernels  and unitarity.

In relation to the amplitudes involving the $\sigma\sigma$ channel, we follow a novel method to calculate them without introducing any new free parameter. The $\sigma$ is a pole due to the $I=0$ S-wave pion interaction \cite{ddecays}. Then, to calculate the elementary 
amplitude $A\to \sigma\sigma$, $N_{A,\sigma\sigma}$, one has to consider first the 
$A\to(\pi\pi)_0(\pi\pi)_0$ tree level amplitude from ${\cal L}_2+{\cal L}_S$, $T_{A}$. 
\footnote{For definiteness, let us consider $A\ne\sigma\sigma$. The method is easily generalized for that case.} The rescattering of the two pairs of pions is taken into account by 
 multiplying $T_{A}$ by $1/(D(s_1)D(s_2))$, where $s_i$ is the total center of mass energy squared of the $i_{\textrm{th}}$ pair  and  $D(s)=1+V_{2}(s)g_{\pi\pi}(s)$ \cite{ddecays}, where 
 $V_2(s) = (s-m_\pi^2)/f^2$ calculated from ${\cal L}_2$. To isolate the $N_{A,\sigma\sigma}$ amplitude, one has to move to the $\sigma$ pole, $s_{\sigma}$, taking the following limit:
\begin{equation}
\lim_{s_1,s_2 \rightarrow s_{\sigma} } \frac{T_A}{D_{II}(s_1)D_{II}(s_2)} = N_{A,\sigma\sigma} \frac{g_{\sigma\pi\pi}^2}{(s_1 - s_\sigma)(s_2 - s_\sigma)}~,
\end{equation}
where  $g_{\sigma\pi\pi}$ is the $\sigma$ coupling to $\pi\pi$ and 
the subscript $II$ indicates that the corresponding function is calculated on the
second Riemann sheet, because it is where the $\sigma$ pole is located. Performing a Laurent expansion of $D_{II}(s)^{-1}$ around $s_\sigma$, $D_{II}(s)^{-1} = \alpha_0/(s-s_\sigma) + \cdots$, the previous limit reduces to $N_{A,\sigma\sigma} = (\alpha_0/g_{\sigma\pi\pi})^2 T_A$. One can show that $(\alpha_0/g_{\sigma\pi\pi})^2 \simeq f^2$ \cite{Albaladejo:2008qa}. Employing $s_i = s_\sigma$ 
to evaluate  $N_{A,\sigma\sigma}$ violates unitarity since then $N_{A,\sigma\sigma}$ would be complex due to the imaginary part of $s_\sigma$. To avoid this point, we interpret the large width of the $\sigma$ as a Lorentzian mass distribution, folding the $\sigma$ masses ($\sqrt{s_{i}}$) used to calculate $N_{A,\sigma\sigma}$ and $g_{\sigma\sigma}(s)$ with that distribution \cite{Albaladejo:2008qa}.

\section{Results and data}

\begin{figure}
{\tiny % GNUPLOT: LaTeX picture with Postscript
\begingroup%
\makeatletter%
\newcommand{\GNUPLOTspecial}{%
  \@sanitize\catcode`\%=14\relax\special}%
\setlength{\unitlength}{0.0500bp}%
\begin{picture}(6479,6048)(0,0)%
  \special{psfile=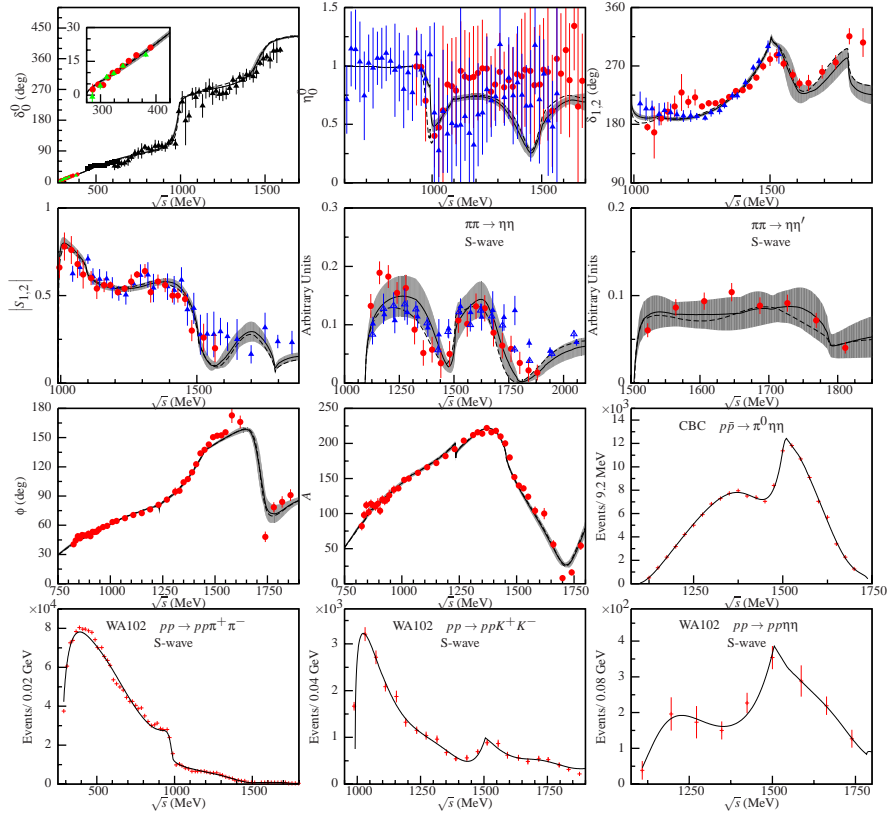 llx=0 lly=0 urx=324 ury=302 rwi=3240}
  \put(5218,1153){\makebox(0,0)[l]{\strut{}S-wave}}%
  \put(4855,1285){\makebox(0,0)[l]{\strut{}WA102\quad $ pp \to p p\eta\eta$}}%
  \put(4265,1417){\makebox(0,0)[l]{\strut{}$\times\! 10^2$}}%
  \put(5400,-55){\makebox(0,0){\strut{}$\sqrt{s}$ (MeV)}}%
  \put(4272,756){%
  \special{ps: gsave currentpoint currentpoint translate
270 rotate neg exch neg exch translate}%
  \makebox(0,0){\strut{}Events/ 0.08 GeV}%
  \special{ps: currentpoint grestore moveto}%
  }%
  \put(6183,35){\makebox(0,0){\strut{} 1750}}%
  \put(5557,35){\makebox(0,0){\strut{} 1500}}%
  \put(4930,35){\makebox(0,0){\strut{} 1250}}%
  \put(4444,1174){\makebox(0,0)[r]{\strut{} 4}}%
  \put(4444,904){\makebox(0,0)[r]{\strut{} 3}}%
  \put(4444,635){\makebox(0,0)[r]{\strut{} 2}}%
  \put(4444,365){\makebox(0,0)[r]{\strut{} 1}}%
  \put(4444,95){\makebox(0,0)[r]{\strut{} 0}}%
  \put(3058,1153){\makebox(0,0)[l]{\strut{}S-wave}}%
  \put(2695,1285){\makebox(0,0)[l]{\strut{}WA102\quad $ pp \to p p K^+ K^-$}}%
  \put(2105,1417){\makebox(0,0)[l]{\strut{}$\times\! 10^3$}}%
  \put(3240,-55){\makebox(0,0){\strut{}$\sqrt{s}$ (MeV)}}%
  \put(2112,756){%
  \special{ps: gsave currentpoint currentpoint translate
270 rotate neg exch neg exch translate}%
  \makebox(0,0){\strut{}Events/ 0.04 GeV}%
  \special{ps: currentpoint grestore moveto}%
  }%
  \put(3861,35){\makebox(0,0){\strut{} 1750}}%
  \put(3383,35){\makebox(0,0){\strut{} 1500}}%
  \put(2905,35){\makebox(0,0){\strut{} 1250}}%
  \put(2428,35){\makebox(0,0){\strut{} 1000}}%
  \put(2284,1153){\makebox(0,0)[r]{\strut{} 3}}%
  \put(2284,800){\makebox(0,0)[r]{\strut{} 2}}%
  \put(2284,448){\makebox(0,0)[r]{\strut{} 1}}%
  \put(2284,95){\makebox(0,0)[r]{\strut{} 0}}%
  \put(898,1153){\makebox(0,0)[l]{\strut{}S-wave}}%
  \put(535,1285){\makebox(0,0)[l]{\strut{}WA102\quad $ pp \to p p\pi^+\pi^-$}}%
  \put(-54,1417){\makebox(0,0)[l]{\strut{}$\times\! 10^4$}}%
  \put(1080,-55){\makebox(0,0){\strut{}$\sqrt{s}$ (MeV)}}%
  \put(-48,756){%
  \special{ps: gsave currentpoint currentpoint translate
270 rotate neg exch neg exch translate}%
  \makebox(0,0){\strut{}Events/ 0.02 GeV}%
  \special{ps: currentpoint grestore moveto}%
  }%
  \put(1637,35){\makebox(0,0){\strut{} 1500}}%
  \put(1051,35){\makebox(0,0){\strut{} 1000}}%
  \put(465,35){\makebox(0,0){\strut{} 500}}%
  \put(124,1270){\makebox(0,0)[r]{\strut{} 8}}%
  \put(124,976){\makebox(0,0)[r]{\strut{} 6}}%
  \put(124,683){\makebox(0,0)[r]{\strut{} 4}}%
  \put(124,389){\makebox(0,0)[r]{\strut{} 2}}%
  \put(124,95){\makebox(0,0)[r]{\strut{} 0}}%
  \put(4855,2797){\makebox(0,0)[l]{\strut{}CBC\quad $ p\bar{p} \to \pi^0\eta\eta$}}%
  \put(4265,2929){\makebox(0,0)[l]{\strut{}$\times\! 10^3$}}%
  \put(5400,1457){\makebox(0,0){\strut{}$\sqrt{s}$ (MeV)}}%
  \put(4272,2268){%
  \special{ps: gsave currentpoint currentpoint translate
270 rotate neg exch neg exch translate}%
  \makebox(0,0){\strut{}Events/ 9.2 MeV}%
  \special{ps: currentpoint grestore moveto}%
  }%
  \put(6308,1547){\makebox(0,0){\strut{} 1750}}%
  \put(5635,1547){\makebox(0,0){\strut{} 1500}}%
  \put(4963,1547){\makebox(0,0){\strut{} 1250}}%
  \put(4444,2841){\makebox(0,0)[r]{\strut{} 14}}%
  \put(4444,2665){\makebox(0,0)[r]{\strut{} 12}}%
  \put(4444,2488){\makebox(0,0)[r]{\strut{} 10}}%
  \put(4444,2312){\makebox(0,0)[r]{\strut{} 8}}%
  \put(4444,2136){\makebox(0,0)[r]{\strut{} 6}}%
  \put(4444,1960){\makebox(0,0)[r]{\strut{} 4}}%
  \put(4444,1783){\makebox(0,0)[r]{\strut{} 2}}%
  \put(4444,1607){\makebox(0,0)[r]{\strut{} 0}}%
  \put(3240,1457){\makebox(0,0){\strut{}$\sqrt{s}$ (MeV)}}%
  \put(2052,2268){%
  \special{ps: gsave currentpoint currentpoint translate
270 rotate neg exch neg exch translate}%
  \makebox(0,0){\strut{}$A$}%
  \special{ps: currentpoint grestore moveto}%
  }%
  \put(4062,1547){\makebox(0,0){\strut{} 1750}}%
  \put(3629,1547){\makebox(0,0){\strut{} 1500}}%
  \put(3197,1547){\makebox(0,0){\strut{} 1250}}%
  \put(2764,1547){\makebox(0,0){\strut{} 1000}}%
  \put(2332,1547){\makebox(0,0){\strut{} 750}}%
  \put(2284,2929){\makebox(0,0)[r]{\strut{} 250}}%
  \put(2284,2665){\makebox(0,0)[r]{\strut{} 200}}%
  \put(2284,2400){\makebox(0,0)[r]{\strut{} 150}}%
  \put(2284,2136){\makebox(0,0)[r]{\strut{} 100}}%
  \put(2284,1871){\makebox(0,0)[r]{\strut{} 50}}%
  \put(2284,1607){\makebox(0,0)[r]{\strut{} 0}}%
  \put(1080,1457){\makebox(0,0){\strut{}$\sqrt{s}$ (MeV)}}%
  \put(-108,2268){%
  \special{ps: gsave currentpoint currentpoint translate
270 rotate neg exch neg exch translate}%
  \makebox(0,0){\strut{}$\phi$ (deg)}%
  \special{ps: currentpoint grestore moveto}%
  }%
  \put(1751,1547){\makebox(0,0){\strut{} 1750}}%
  \put(1356,1547){\makebox(0,0){\strut{} 1500}}%
  \put(962,1547){\makebox(0,0){\strut{} 1250}}%
  \put(567,1547){\makebox(0,0){\strut{} 1000}}%
  \put(172,1547){\makebox(0,0){\strut{} 750}}%
  \put(124,2929){\makebox(0,0)[r]{\strut{} 180}}%
  \put(124,2709){\makebox(0,0)[r]{\strut{} 150}}%
  \put(124,2488){\makebox(0,0)[r]{\strut{} 120}}%
  \put(124,2268){\makebox(0,0)[r]{\strut{} 90}}%
  \put(124,2048){\makebox(0,0)[r]{\strut{} 60}}%
  \put(124,1827){\makebox(0,0)[r]{\strut{} 30}}%
  \put(124,1607){\makebox(0,0)[r]{\strut{} 0}}%
  \put(5400,4177){\makebox(0,0)[l]{\strut{}S-wave}}%
  \put(5400,4309){\makebox(0,0)[l]{\strut{}$\pi\pi \to \eta\eta^\prime$}}%
  \put(5400,2969){\makebox(0,0){\strut{}$\sqrt{s}$ (MeV)}}%
  \put(4212,3780){%
  \special{ps: gsave currentpoint currentpoint translate
270 rotate neg exch neg exch translate}%
  \makebox(0,0){\strut{}Arbitrary Units}%
  \special{ps: currentpoint grestore moveto}%
  }%
  \put(6049,3059){\makebox(0,0){\strut{} 1800}}%
  \put(5530,3059){\makebox(0,0){\strut{} 1700}}%
  \put(5011,3059){\makebox(0,0){\strut{} 1600}}%
  \put(4492,3059){\makebox(0,0){\strut{} 1500}}%
  \put(4444,4441){\makebox(0,0)[r]{\strut{} 0.2}}%
  \put(4444,3780){\makebox(0,0)[r]{\strut{} 0.1}}%
  \put(4444,3119){\makebox(0,0)[r]{\strut{} 0}}%
  \put(3240,4177){\makebox(0,0)[l]{\strut{}S-wave}}%
  \put(3240,4309){\makebox(0,0)[l]{\strut{}$\pi\pi \to \eta\eta$}}%
  \put(3240,2969){\makebox(0,0){\strut{}$\sqrt{s}$ (MeV)}}%
  \put(2052,3780){%
  \special{ps: gsave currentpoint currentpoint translate
270 rotate neg exch neg exch translate}%
  \makebox(0,0){\strut{}Arbitrary Units}%
  \special{ps: currentpoint grestore moveto}%
  }%
  \put(3983,3059){\makebox(0,0){\strut{} 2000}}%
  \put(3570,3059){\makebox(0,0){\strut{} 1750}}%
  \put(3157,3059){\makebox(0,0){\strut{} 1500}}%
  \put(2745,3059){\makebox(0,0){\strut{} 1250}}%
  \put(2332,3059){\makebox(0,0){\strut{} 1000}}%
  \put(2284,4441){\makebox(0,0)[r]{\strut{} 0.3}}%
  \put(2284,4000){\makebox(0,0)[r]{\strut{} 0.2}}%
  \put(2284,3560){\makebox(0,0)[r]{\strut{} 0.1}}%
  \put(2284,3119){\makebox(0,0)[r]{\strut{} 0}}%
  \put(1080,2969){\makebox(0,0){\strut{}$\sqrt{s}$ (MeV)}}%
  \put(-108,3780){%
  \special{ps: gsave currentpoint currentpoint translate
270 rotate neg exch neg exch translate}%
  \makebox(0,0){\strut{}$\left| S_{1,2} \right|$}%
  \special{ps: currentpoint grestore moveto}%
  }%
  \put(1219,3059){\makebox(0,0){\strut{} 1500}}%
  \put(193,3059){\makebox(0,0){\strut{} 1000}}%
  \put(124,4441){\makebox(0,0)[r]{\strut{} 1}}%
  \put(124,3780){\makebox(0,0)[r]{\strut{} 0.5}}%
  \put(124,3119){\makebox(0,0)[r]{\strut{} 0}}%
  \put(5400,4481){\makebox(0,0){\strut{}$\sqrt{s}$ (MeV)}}%
  \put(4212,5292){%
  \special{ps: gsave currentpoint currentpoint translate
270 rotate neg exch neg exch translate}%
  \makebox(0,0){\strut{}$\delta_{1,2}$ (deg)}%
  \special{ps: currentpoint grestore moveto}%
  }%
  \put(5539,4571){\makebox(0,0){\strut{} 1500}}%
  \put(4513,4571){\makebox(0,0){\strut{} 1000}}%
  \put(4444,5953){\makebox(0,0)[r]{\strut{} 360}}%
  \put(4444,5512){\makebox(0,0)[r]{\strut{} 270}}%
  \put(4444,5072){\makebox(0,0)[r]{\strut{} 180}}%
  \put(4444,4631){\makebox(0,0)[r]{\strut{} 90}}%
  \put(3240,4481){\makebox(0,0){\strut{}$\sqrt{s}$ (MeV)}}%
  \put(2052,5292){%
  \special{ps: gsave currentpoint currentpoint translate
270 rotate neg exch neg exch translate}%
  \makebox(0,0){\strut{}$ \eta_{0}^{0}$}%
  \special{ps: currentpoint grestore moveto}%
  }%
  \put(3818,4571){\makebox(0,0){\strut{} 1500}}%
  \put(2992,4571){\makebox(0,0){\strut{} 1000}}%
  \put(2284,5953){\makebox(0,0)[r]{\strut{} 1.5}}%
  \put(2284,5512){\makebox(0,0)[r]{\strut{} 1}}%
  \put(2284,5072){\makebox(0,0)[r]{\strut{} 0.5}}%
  \put(2284,4631){\makebox(0,0)[r]{\strut{} 0}}%
  \put(911,5176){\makebox(0,0){\strut{} 400}}%
  \put(493,5176){\makebox(0,0){\strut{} 300}}%
  \put(340,5802){\makebox(0,0)[r]{\strut{} 30}}%
  \put(340,5547){\makebox(0,0)[r]{\strut{} 15}}%
  \put(340,5292){\makebox(0,0)[r]{\strut{} 0}}%
  \put(1080,4481){\makebox(0,0){\strut{}$\sqrt{s}$ (MeV)}}%
  \put(-108,5292){%
  \special{ps: gsave currentpoint currentpoint translate
270 rotate neg exch neg exch translate}%
  \makebox(0,0){\strut{}$\delta_{0}^{0}$ (deg)}%
  \special{ps: currentpoint grestore moveto}%
  }%
  \put(1733,4571){\makebox(0,0){\strut{} 1500}}%
  \put(1096,4571){\makebox(0,0){\strut{} 1000}}%
  \put(459,4571){\makebox(0,0){\strut{} 500}}%
  \put(124,5798){\makebox(0,0)[r]{\strut{} 450}}%
  \put(124,5567){\makebox(0,0)[r]{\strut{} 360}}%
  \put(124,5335){\makebox(0,0)[r]{\strut{} 270}}%
  \put(124,5103){\makebox(0,0)[r]{\strut{} 180}}%
  \put(124,4871){\makebox(0,0)[r]{\strut{} 90}}%
  \put(124,4639){\makebox(0,0)[r]{\strut{} 0}}%
\end{picture}%
\endgroup
  }
\caption{From left to right and top to bottom: $S$-wave $\pi\pi$ phase shift $\delta_0^0$,  elasticity parameter $\eta_0^0$, phase of the $\pi\pi\to K\bar{K}$ $S$-wave $\delta_{1,2}$,
its modulus  $|S_{12}|$, $S$-wave event distributions to the $\pi\pi\to\eta\eta$ and $\eta\eta'$ 
reactions and  the phase ($\phi$) and  modulus ($A$) of $K^-\pi^+\to K^-\pi^+$. The last four panels correspond to the 
Crystal Barrel and WA102 Collaborations on $p\bar{p}$ annihilation and $pp$ central production, respectively.\label{fig}}
\end{figure}

From the $T$-matrix, we can calculate the $S$-matrix elements, $S_{i,j} = \delta_{i,j} + 2i T_{i,j}\sqrt{q_i q_j}/8\pi\sqrt{s}$, with $q_i$ the centre of mass three-momentum of channel $i$. The free parameters in our theory are the subtraction constants $a_i$ in the functions $g_i(s)$ and the masses and coupling constants involved in $\mathcal{L}_S$. The number of \textit{free} subtraction constants is reduced because  we take $a_{\rho\rho} = a_{\omega\omega} = a_{K^*\bar{K}^*} = a_{\omega\phi} = a_{\phi\phi}$, since $SU(3)$ breaking is milder in the vector sector. We can also fix some of the parameters related to the bare resonances 
required by our fits,  two octets and one singlet. Namely, from ref. \cite{Jamin:2000wn} the first octet is set to $M_8^{(1)} = 1.29\ \textrm{GeV}$, $c_d^{(1)} = c_m^{(1)} = 26\ \textrm{MeV}$. The  mass of the second octet is also fixed, $M_8^{(2)} = 1.90\ \textrm{GeV}$ from the same reference. So we are left with 3 parameters for the singlet and 2 for the second octet, plus 7 free subtraction constants, totaling 12 free parameters to fit 370 the experimental data of the first six panels in fig.\ref{fig}, from left to right and top to bottom. The data of the last four panels, in the same order, were fitted 
as a sum of Breit-Wigner's plus a soft background, with the values used
for the pole positions and strong couplings to the final states  given by the previous fit. In these last data one can clearly observe peaks corresponding 
to the $f_0(1500)$, $f_0(980)$ and $\sigma$. The $f_0(1710)$ is also needed 
to reproduce the shoulders above 1.5~GeV in several reactions. 
See ref.\cite{Albaladejo:2008qa} for further 
details. One observes a good reproduction of the data. Compared with previous works, we have fewer free parameters to reproduce more data, and this can be done because we determine the interaction kernels from chiral Lagrangians which allows us to include many more channels and to avoid ad-hoc parametrizations.

\section{Spectroscopy}

\begin{table}
\caption{List of the poles found on the different Riemann sheets and
couplings  of the $f_0(1370)$, $f_0^R$ and $f_0(1710)$. Some branching ratios for the $f_0(1710)$ are also shown.\label{ta-poles}}
\begin{tabular}{cc}
\begin{tabular}{ccc}
\multicolumn{3}{c}{$I=0$} \\ \hline
Pole & $\textrm{Re}{\sqrt{s}}$ & $\textrm{Im}{\sqrt{s}}$ \\ \hline
$f_0(600) = \sigma$ & $456 \pm 6$ & $241 \pm 7$ \\
$f_0(980)$ & $983 \pm 4$ & $25 \pm 3$ \\
$f_0^L$ & $1466 \pm 15$ & $158 \pm 12$ \\
$f_0^R$ & $1602 \pm 15$ & $44 \pm 15$ \\
$f_0(1710)$ & $1690 \pm 20$ & $110 \pm 20$ \\
$f_0(1790)$ & $1810 \pm 15$ & $190 \pm 20$ \\ \hline
\end{tabular}
&
\begin{tabular}{ccc}
\multicolumn{3}{c}{$I=1/2$} \\ \hline
Pole & $\textrm{Re}{\sqrt{s}}$ & $\textrm{Im}{\sqrt{s}}$ \\ \hline
$K^*_0(800) = \kappa$ & $708 \pm 6$ & $313 \pm 10$ \\
$K^*_0(1430)$ & $1435 \pm 6$ & $142 \pm 8$ \\
$K^*_0(1950)$ & $1750 \pm 20$ & $150 \pm 20$ \\
  \hline
\end{tabular}
\\
\begin{tabular}{cccc}
\multicolumn{4}{c}{Couplings} \\ \hline
    GeV               &	 $f_0(1370)$     & $f_0^R$        & $f_0(1710)$\\ \hline
$|g_{\pi^+\pi^-}|$      & $3.59\pm 0.16$ & $1.31\pm 0.22$ & $1.24\pm 0.16$   \\
$|g_{K^0{\bar{K}}^0}|$  & $2.23\pm 0.18$ & $2.06\pm 0.17$ & $2.0\pm 0.3$     \\ 
$|g_{\eta\eta}|$        & $1.7\pm 0.3$   & $3.78\pm 0.26$ & $3.3\pm 0.8$     \\
$|g_{\eta\eta'}|$       & $4.0\pm 0.3$   & $4.99\pm 0.24$ & $5.1\pm 0.8$     \\
$|g_{\eta'\eta'}|$      & $3.7\pm 0.4$   & $8.3\pm 0.6$   & $11.7\pm 1.6$    \\ \hline
\end{tabular}
&
\begin{tabular}{ccc}
\multicolumn{3}{c}{Branching ratios of $f_0(1710)$} \\ \hline
 & Value & PDG \\ \hline
 $\frac{\Gamma(K\bar{K})}{\Gamma(\textrm{total})}$ & $0.36 \pm 0.12$ & $0.38^{+0.09}_{-0.19}$ \\
$\frac{\Gamma(\eta\bar{\eta})}{\Gamma(\textrm{total})}$ & $0.22 \pm 0.12$ & $0.18^{+0.03}_{-0.13}$ \\
$\frac{\Gamma(\pi\pi)}{\Gamma(K\bar{K})}$ & $0.32 \pm 0.14$ & $<0.11$ \\ \hline
\end{tabular}
\end{tabular}
\end{table}

Moving to the complex plane, we find the poles given in table \ref{ta-poles}. For $I=1/2$, we have reproduced the resonances found in the PDG. For $I=0$, we have the $\sigma$, $f_0(980)$, $f_0(1370)$, $f_0(1500)$, $f_0(1710)$ and $f_0(1790)$ resonances. The width of $f_0(1710)$ from PDG is $137 \pm 8\ \textrm{MeV}$, which is smaller than $220 \pm 40\ \textrm{MeV}$, as determined from pole position. However, on the real axis, the value of the width corresponding to the half-maximum for the partial waves with $f_0(1710)$ signals is $160\ \textrm{MeV}$ \cite{Albaladejo:2008qa}, recovering in that way the agreement with PDG. This reduction is due to the opening of several channels along the resonance region. Our determination agrees with the parameters reported by BESII for the $f_0(1790)$. The explanation for $f_0(1370)$ and $f_0(1500)$ is more complicated. The $f_0(1370)$ is mainly given by the pole $f_0^L$, though the precise shape of the amplitudes on the real axis is sensitive to the $f_0^R$ pole  for some channels. This last pole is located on a Riemann sheet which does not influence \textit{directly} the real axis beyond the $\eta\eta'$ threshold, at $\sqrt{s} = 1505\ \textrm{MeV}$. This effect typically gives raise to a pronounced signal at the threshold, and that is the reason to have the mass of the $f_0(1500)$ at $1505 \pm 6\ \textrm{MeV}$. From the pole position, one could think that the width is $88\ \textrm{MeV}$. However, given a Breit-Wigner located at the position of the $f_0^R$ pole, the energy interval below $1.5\ \textrm{GeV}$ at which half the value of the amplitude squared at the maximum (at $1.5\ \textrm{GeV}$) is reached is $\delta = 1.2 \Gamma = 105\ \textrm{MeV}$, which is, not by chance, the width of the $f_0(1500)$.
%(2\textrm{Im}\sqrt{s}_{f_0^R})

Consider now the couplings given in table \ref{ta-poles}. The ones of $f_0^L-f_0(1370)$ correspond to the pure $I=0$ octet member, because they are very close to the bare octet ones \cite{Albaladejo:2008qa}, calculated from 
${\cal L}_S$ with $M_8^{(1)}$, $c_d^{(1)}$ and $c_m^{(1)}$ fixed above. This is also the case for the $K_0^*(1430)$ resonance, which is the $I=1/2$ member of the same octet. So the first octet is a pure one, without mixing with the $f_0^R$ nor $f_0(1710)$. In addition, these couplings imply a large $\pi\pi$ width $\Gamma(f_0(1370) \to 4\pi) / \Gamma(f_0(1370) \to 2\pi) = 0.30 \pm 0.12$, in agreement with the recent determination of ref. \cite{Bugg:2007ja}. The couplings for $f_0(1710)$ and $f_0^R$ are similar, due to the fact that these poles move into each other in a continuous transition between 
 their respective Riemann sheets. From the couplings of the $f_0(1710)$ we can calculate some branching ratios, given in table \ref{ta-poles}, together with the values of the PDG, and they are compatible within one sigma.  Finally, we obtain that the $f_0(1790)$ has a small $K\bar{K}$ coupling, a major difference with respect to the $f_0(1710)$ as stressed by BESII.

Let us now see that the pattern of the couplings of the $f_0^R$ and the $f_0(1710)$ corresponds to the chiral suppression mechanism of the coupling of a scalar glueball to $\bar{q}q$ \cite{Chanowitz:2005du}.
 This mechanism predicts that this coupling is proportional to the quark mass, so  $\bar{u}u$ or $\bar{d}d$ production is strongly suppressed compared with $\bar{s}s$. With an $\eta$--$\eta'$ mixing angle $\sin\theta = -1/3$, one has that $\eta = -\eta_s/\sqrt{3} + \eta_{n}\sqrt{2/3}$ and $\eta' = \eta_{s}\sqrt{2/3} + \eta_{n}/\sqrt{3}$, where $\eta_{s} = \bar{s}s$ and $\eta_{n} = (\bar{u}u + \bar{d}d)/\sqrt{2}$. Denoting  by $g_{ss}$, $g_{ns}$ and $g_{nn}$ the production of $\eta_{s}\eta_{s}$, $\eta_{n}\eta_{s}$ and $\eta_{n}\eta_{n}$, in order, one has 
$g_{\eta'\eta'}=2g_{ss}/3+g_{nn}/3+2\sqrt{2}g_{ns}/3~,~
g_{\eta\eta'}=-\sqrt{2}g_{ss}/3+\sqrt{2}g_{nn}/3+g_{ns}/3~,~
g_{\eta\eta}=g_{ss}/3+2g_{nn}/3-2\sqrt{2}g_{ns}/3.$ Taking  into account  the numerical values given in table \ref{ta-poles} for the couplings 
of the $f_0^R$  we then 
find  $g_{ss} = 11.5 \pm 0.5$, $g_{ns} = -0.2$ and $g_{nn} = -1.4\ \textrm{GeV}$, and the suppression is clear. Consider now the $K^0\bar{K}^0$, where $K^0 = \sum_{i=1}^3 \bar{s}_i u^i/\sqrt{3}$, summing over the color indices. The production of a colour singlet $\bar{s}s$ from $\bar{K}^0 K^0$ requires the combination $\bar{s}_i s^j = \delta_{i}^{j} \bar{s}s/3 + (\bar{s}_i s^j - \delta_{i}^{j} \bar{s}s/3)$, and similarly for $\bar{u}_j u^i$. As only the configuration $\bar{s}s\bar{u}u$ contributes, it picks a factor 1/3. In addition, $g_{ss}$ takes an extra factor 2 compared to $\bar{s}s\bar{u}u$, because the former contains two $\bar{s}s$. One then expects the coupling of $K^0\bar{K}^0$ to have the absolute value $g_{ss}/6$, as it is the case for $f_0^R$ and $f_0(1710)$. Also, quenched lattice QCD \cite{qcd} agrees with the fact that the coupling of the lightest scalar glueball to pseudoscalar pairs in the $SU(3)$ limit scales as the quark mass, supporting the chiral suppression mechanism, as our results does. This mechanism also implies that the glueball should remain unmixed, which fits with our statement that the $f_0(1710)$ and $f_0^R$ does not mix with $f_0^L$. In addition, the masses of $f_0(1710)$ and $f_0^R$ agree with the quenched lattice QCD prediction for the lightest scalar glueball, $(1.66 \pm 0.05)\ \textrm{GeV}$.

In summary, we have presented a detailed coupled channel study of the $I=0,1/2$ meson-mesons $S$-waves up to $2\ \textrm{GeV}$, including the necessary channels, reproducing the $0^{++}$ and $1/2^{++}$ resonances below that energy. We have identified the $f_0(1710)$ and $f_0^R$ pole (which is an \textit{important contribution} to the $f_0(1500)$) as glueballs. The pole $f_0^L$, the main contribution to $f_0(1370)$, turns out to be a pure octet member.

\end{document}